\documentclass[12pt,epsf]{article}
\usepackage{amsfonts}
\usepackage{amscd,amssymb,amsmath,epsf}

\advance \topmargin by -\headheight
\advance \topmargin by -\headsep
\evensidemargin \oddsidemargin
\begin{document}

%%%%%%%%%%%%%%%%%%%%%%%%%%%%%%%%%%%%%%%%%%%%%%%%%%%%%%%%%%%%%%%%%%%%%%%%%%%%%%%%
\begin{titlepage}
%%%%%%%%%%%%%%%%%%%%%%%%%%%%%%%%%%%%%%%%%%%%%%%%%%%%%%%%%%%%%%%%%%%%%%%%%%%%%%%%

\begin{center} \Large \bf Lie--algebra expansions, Chern--Simons theories and the
Einstein--Hilbert lagrangian
\end{center}

\vskip 0.3truein
\begin{center}
Jos\'e D. Edelstein${}^{\,\star\natural}$ \footnote{jedels-at-usc.es}, Mokhtar
Hassa\"{\i}ne${}^{\,\ddagger\natural}$
\footnote{hassaine-at-inst-mat.utalca.cl}, Ricardo Troncoso${}^{\,\natural}$
\footnote{ratron-at-cecs.cl} and Jorge Zanelli${}^{\,\natural}$
\footnote{jz-at-cecs.cl}

\vspace{0.3in}

${}^{\,\star}$Departamento de F\'\i sica de Part\'\i culas, Universidade de
Santiago de Compostela \\ and Instituto Galego de F\'\i sica de Altas
Enerx\'\i as (IGFAE) \\ E-15782 Santiago de Compostela, Spain
%\vspace{0.2in}

${}^{\,\natural}$Centro de Estudios Cient\'\i ficos (CECS) Casilla 1469,
Valdivia, Chile
%\vspace{0.2in}

${}^{\,\ddagger}$Instituto de Matem\'atica y F\'{\i}sica, Universidad
de Talca, Casilla 747, Talca, Chile

\end{center}
%\vskip.5truein

%%%%%%%%%%%%%%%%%%%%%%%%%%%%%%%%%%%%%%%%%%%%%%%%%%%%%%%%%%%%%%%%%%%%%%%%%%%%%%%%
\begin{center}
\bf ABSTRACT
\end{center}
Starting from gravity as a Chern--Simons action for the AdS algebra in five
dimensions, it is possible to deform the theory through an expansion of the Lie
algebra that leads to a system consisting of the Einstein--Hilbert action plus
nonminimally coupled matter. The deformed system is gauge invariant under the
Poincar\'e group enlarged by an Abelian ideal. Although the resulting action
naively looks like General Relativity plus corrections due to matter sources,
it is shown that the nonminimal couplings produce a radical departure from GR.
Indeed, the dynamics is not continuously connected to the one obtained from
Einstein--Hilbert action. In a matter--free configuration and in the
torsionless sector, the field equations are too strong a restriction on the
geometry as the metric must satisfy both the Einstein and pure Gauss--Bonnet
equations. In particular, the five-dimensional Schwarzschild geometry fails to
be a solution; however, configurations corresponding to a brane-world with
positive cosmological constant on the worldsheet are admissible when one of the
matter fields is switched on. These results can be extended to higher odd
dimensions.
%%%%%%%%%%%%%%%%%%%%%%%%%%%%%%%%%%%%%%%%%%%%%%%%%%%%%%%%%%%%%%%%%%%%%%%%%%%%%%%%

\vskip2.3truecm \leftline{CECS-PHY-06/09} \leftline{hep-th/0605174 \hfill May
2006}
\smallskip
\end{titlepage}\setcounter{footnote}{0}

%%%%%%%%%%%%%%%%%%%%%%%%%%%%%%%%%%%%%%%%%%%%%%%%%%%%%%%%%%%%%%%%%%%%%%%%%%%%%%%%

\section{Introduction}

%%%%%%%%%%%%%%%%%%%%%%%%%%%%%%%%%%%%%%%%%%%%%%%%%%%%%%%%%%%%%%%%%%%%%%%%%%%%%%%%

Field theories in higher dimensions are nowadays a natural part of the high
energy physics toolkit. Having accepted the possibility of higher dimensions
makes it worthwhile to consider theories with structures that are inherently
higher dimensional, and not just a lifting of those that exist in four
dimensions. In particular, gravitation theories which are not just the
dimensional continuation of the Einstein--Hilbert action deserve special
attention. This is the case of the Gauss--Bonnet action, which only exists for
$d\geq5$ dimensions, and shares the basic assumptions of General Relativity,
namely, general covariance and second order field equations for the metric. In
general, for $d>4$, the generic theory that accomplishes these criteria has
higher powers of curvature, and is given by the Lovelock action \cite{Lov}.

As a purely metric theory, the Lovelock action describes the same number of
degrees of freedom as general relativity (GR) \cite{TeZ}. However, in the first
order formalism, where the vielbein and the spin connection are independently
varied, there could be extra degrees of freedom associated to the fact that the
torsion may not vanish in vacuum \cite{TrZcqg}, unlike in GR. This makes it
natural to introduce extra terms containing torsion in the lagrangian
\cite{MZ}. It is also natural to wonder whether these theories admit a locally
supersymmetric extension. This was shown to be possible for a very particular
case of Einstein--Gauss--Bonnet gravity in five dimensions \cite{Cham}, where
the coupling of the Gauss--Bonnet term is chosen so that the lagrangian can be
viewed as a Chern--Simons (CS) form for the super AdS$_{5}$ group. It has also
been recently shown that Euclidean gravity as a CS theory in five dimensions
can be seen to emerge from a CS\ theory in $0+1$ dimensions in the large $N$
limit \cite{Nair}.

The supersymmetric extension of the lagrangian defined purely by the term with
the highest power of curvature in any odd dimension, can also be seen as a CS
form for the super Poincar\'e algebra with a ``central extension"
\cite{BTrZ,HOT,HTZ}.

The supersymmetric extensions of the Lovelock theories for higher odd
dimensions with a negative cosmological constant can be achieved provided the
gravitational sector is supplemented with appropriate torsion terms so that the
entire action can be viewed as a CS form for the minimal super AdS group
\cite{TrZsugra,JZrio}.

It is worth highlighting that in eleven-dimensions, apart from the standard
Cremmer--Julia--Scherk supergravity theory \cite{CJS}, there exists an AdS
supergravity whose gravitational sector has higher powers in the curvature. As
pointed out in Ref. \cite{TrZsugra} this theory has $OSp(1|32)$ gauge
invariance, and some sectors of it might be related to the low energy limit of
$M$--theory if one identifies the totally antisymmetric part of the contorsion
with the 3-form. Its dual 6-form is also present, and hence, the theory not
only has the potential to contain standard supergravity, but also some kind of
dual version of it.

This suggestion has been further developed in \cite{Horava}, where it is
proposed that M--theory could be related to a Chern--Simons theory with an
enlarged gauge group, namely, $OSp(1|32) \times OSp(1|32)$. However, to this
date, the contact with the Cremmer--Julia--Scherk theory remains elusive
(see also \cite{MaxB,Nastase}).

On the other hand, a general method to expand a Lie (super) algebra allows to
consistently deform a Chern--Simons theory with a given Lie algebra into
another one whose Lie algebra has more generators \cite{AIPV}. It is simple to
see that applying this method to the eleven-dimensional AdS supergravity yields
a deformed theory whose gravitational sector contains the Einstein--Hilbert
action nonminimally coupled to a host of additional matter fields. This
proposal has been explicitly carried out in \cite{Izq}, where it was found that
the field equations of the deformed theory in vacuum do not reduce to the
Einstein field equations, but nevertheless it was shown that the $M$-waves of
standard eleven-dimensional supergravity are also solutions of the new theory.

The purpose of this note is to capture the source of the clash between the
deformed theory and GR in a simplified setting. We consider gravity as a CS
action for the AdS algebra in five dimensions, which is the simplest
non-trivial case where the problem arises. Deforming the theory through an
expansion of the Lie algebra leads to a CS system that is gauge invariant under
the Poincar\'e group with an Abelian ideal. The resulting action consists of
the Einstein--Hilbert term plus other terms containing matter nonminimally
coupled to the curvature. These extra terms do not correspond to corrections of
GR since although the action reduces to Einstein--Hilbert when the matter
fields are switched off, the field equations don't. This is a generic feature
of nonminimal couplings, which produce strong deviations from GR. Indeed, in
the torsionless sector in vacuum, the geometry must satisfy both the Einstein
and pure Gauss--Bonnet equations simultaneously. These restrictions are so
strong as to rule out, for instance, the five-dimensional Schwarzschild
solution, but not a $pp$-wave.

When matter fields are switched on, however, configurations corresponding to
a brane-world with positive cosmological constant on the worldsheet are
admissible. Curiously, the metric for this class of solutions also solves
the field equations of the supergravity theories with local Poincar\'e
invariance in vacuum, as discussed in \cite{HTZ}.

In dimensions $d>5$, the application of the expansion method as in \cite{AIPV}
gives rise to an extension of the Poincar\'e algebra by a large number of extra
generators, which no longer form an Abelian ideal. As a consequence, in the
torsionless vacuum sector, in addition to the Einstein and pure Gauss--Bonnet
equations, all the possible field equations coming from each term in the
Lovelock action must be satisfied. This means that, as the dimensions increase,
the clash with General Relativity becomes unsurmountable. However, in all cases
$pp$-waves are always solutions of the deformed systems.

\section{Five-dimensional Chern--Simons AdS gravity and its expansion}

Nonabelian Chern-Simons theories in five dimensions are given by a
Lagrangian $L$ such that
\begin{equation}  \label{CS5}
d L=\left< \mathbf{F}\wedge \mathbf{F}\wedge \mathbf{F} \right> ~,
\end{equation}
where $\mathbf{F= dA + A} \wedge \mathbf{A}$ is the field strength
(curvature) and $\mathbf{A}$ is the nonabelian gauge field. Here the bracket
$\left< \cdots \right>$ stands for a third rank invariant tensor of the Lie
algebra. For the five-dimensional AdS algebra the connection reads
\begin{equation}
\mathbf{A} = e^a\, J_a + \frac{1}{2}\,\omega^{ab}\, J_{ab} ~,  \label{PoiCon}
\end{equation}
where $J_{ab}$ and $J_{a}$ are the generators of Lorentz transformations and
AdS boosts, respectively, $e^a$ is the vielbein and $\omega^{ab}$ is the spin
connection. The only nonvanishing components of the bracket are
\begin{eqnarray}
\left< J_{ab},J_{cd},J_{e} \right> = \frac{4}{3}\epsilon_{a b c d e}~.
\label{brack2n+1}
\end{eqnarray}
Thus, the action is the Einstein--Gauss--Bonnet theory with fixed relative
couplings, up to surface terms,
\begin{equation}
I = \kappa \int \epsilon_{abcdf}\, \Big( R^{ab} \wedge R^{cd} \wedge e^f +
\frac{2}{3} R^{ab} \wedge e^c \wedge e^d \wedge e^f + \frac{1}{5} e^a\wedge
e^b \wedge e^c\wedge e^d \wedge e^f \Big) ~,  \label{action5}
\end{equation}
where the AdS radius has been fixed to one, and
$R^{ab}=d\omega^{ab}+\omega^a_c\wedge\omega^{cb}$ is the curvature two-form.

Applying the expansion method of \cite{AIPV} up to second order, the
connection is deformed as
\begin{equation}
\mathbf{A}\rightarrow \widetilde{\mathbf{A}}=e^{a}\,P_{a}+\frac{1}{2}%
\,\omega ^{ab}\,J_{ab}+h^{a}\,Z_{a}+\frac{1}{2}\,\kappa ^{ab}\,Z_{ab}~.
\label{A5}
\end{equation}%
This connection is the gauge field for an extension of the Poincar\'{e}
algebra by an Abelian ideal spanned by $\{Z_{a},Z_{ab}\}$, so that the
commutation relations of the full algebra are given by
\begin{eqnarray}
&&\left[ P_{a},P_{b}\right] =Z_{ab}~,\qquad \left[ J_{ab},P_{c}\right]
=P_{a}\,\eta _{bc}-P_{b}\,\eta _{ac}~,  \nonumber  \label{algebra5} \\
&&\left[ J_{ab},J_{cd}\right] =-J_{ac}\,\eta _{bd}+J_{bc}\,\eta
_{ad}-J_{bd}\,\eta _{ac}+J_{ad}\,\eta _{bc}~,  \label{algebrax} \\
&&\left[ Z_{a},Z_{b}\right] =\left[ Z_{ab},Z_{c}\right] =\left[ Z_{ab},Z_{cd}%
\right] =\left[ P_{a},Z_{b}\right] =0~,  \nonumber \\
&&\left[ Z_{ab},P_{c}\right] =\left[ J_{ab},Z_{c}\right] =Z_{a}\,\eta
_{bc}-Z_{b}\,\eta _{ac}~,  \nonumber \\
&&\left[ J_{ab},Z_{cd}\right] =-Z_{ac}\,\eta _{bd}+Z_{bc}\,\eta
_{ad}-Z_{bd}\,\eta _{ac}+Z_{ad}\,\eta _{bc}~.  \nonumber
\end{eqnarray}%
The curvature two-form corresponding to the connection (\ref{A5}), is obtained
as
\begin{eqnarray}
\mathbf{F}\rightarrow \widetilde{\mathbf{F}} &=&T^{a}\,P_{a}+\frac{1}{2}%
\,R^{ab}\,J_{ab}+\left[ Dh^{a}+\kappa _{b}^{\;a}\wedge e^{b}\right] \,Z_{a}+%
\frac{1}{2}\left[ D\kappa ^{ab}+e^{a}\wedge e^{b}\right] Z_{ab}~,  \nonumber
\\
&=&T^{a}\,P_{a}+\frac{1}{2}\,R^{ab}\,J_{ab}+\tilde{F}^{a}\,Z_{a}+\frac{1}{2}%
\,\tilde{F}^{ab}\,Z_{ab}~,  \label{curvature}
\end{eqnarray}%
where $D$ is the covariant derivative with respect to the Lorentz piece of
the connection, and $T^{a}=De^{a}$. The expansion method ensures that one
can consistently isolate the second order term in the expansion of the
Lagragian, which corresponds to a CS theory for the deformed algebra, \emph{%
i. e.},
\begin{equation}
d\widetilde{L}=\left\langle \widetilde{\mathbf{F}}\wedge \widetilde{\mathbf{F%
}}\wedge \widetilde{\mathbf{F}}\right\rangle ~,  \label{CS5s}
\end{equation}%
where now the bracket is defined as
\begin{equation}
\left\langle J_{ab},J_{cd},Z_{f}\right\rangle =\left\langle
J_{ab},Z_{cd},P_{f}\right\rangle =\frac{4}{3}\,\epsilon _{abcdf}~.
\label{brackets5}
\end{equation}

The resulting deformed action is locally invariant under gauge
transformations generated by the algebra (\ref{algebrax}). If expressed in
terms of the components of the gauge field the action reads (we put $\kappa
= 1$),
\begin{equation}
\widetilde{I}= \int \epsilon_{abcdf}\, \Big( \frac{2}{3} R^{ab} \wedge e^c
\wedge e^d \wedge e^f + R^{ab} \wedge R^{cd} \wedge h^f + 2 R^{ab} \wedge
\kappa^{cd} \wedge T^f \Big) ~,  \label{action5'}
\end{equation}
up to surface terms.

A shortcut to obtain the deformed action from (\ref{action5}) is as follows:
shift the fields as
\begin{eqnarray}  \label{spin}
\omega^{ab} \to \omega^{ab} + \ell^{-2}\; \kappa^{ab} ~, \\
e^{a}\to \ell^{-1}\; e^a + \ell^{-3}\; h^{a} ~.  \label{vielbein}
\end{eqnarray}
The deformed action (\ref{action5'}) is then obtained replacing the shifted
fields in the original action (\ref{action5}) and retaining the terms of
order $\ell^{-3}$. As shown in the Appendix, it is straightforward to
generalize this shortcut to apply the expansion procedure in higher
dimensions.

It is apparent from the action (\ref{action5'}) that if one identifies the
field $e^a$ with the vielbein, the system consists of the Einstein--Hilbert
action plus nonminimally coupled matter fields given by $h^a$ and $%
\kappa^{ab}$. Although the resulting action naively looks like general
relativity plus corrections due to the matter sources, one can see that the
nonminimal couplings are so intricate as to produce a radical departure from
general relativity. These extra terms do not correspond to corrections of
General Relativity since although the action reduces to Einstein--Hilbert
when the matter fields are switched off, the field equations don't. Indeed,
varying the action (\ref{action5}) with respect to the vielbein, the spin
connection and the additional bosonic fields, $h^a$ and $\kappa^{ab}$, the
field equations read respectively,
\begin{eqnarray}
& & \epsilon_{abcdf}\, R^{ab} \wedge e^c \wedge e^d = - \epsilon_{abcdf}\,
R^{ab} \wedge D \kappa^{cd} ~,  \label{eq1} \\
& & \epsilon_{abcdf}\, \Big( \widetilde{F}^{cd} \wedge T^f + R^{cd} \wedge
\widetilde{F}^f\Big) = 0 ~,  \label{eq2} \\
& & \epsilon_{abcdf}\, R^{ab} \wedge R^{cd} = 0 ~,  \label{eq3} \\
& & \epsilon_{abcdf}\, R^{ab} \wedge T^c = 0 ~.  \label{eq4}
\end{eqnarray}

In a matter--free configuration and in the torsionless sector, the metric
must satisfy simultaneously the Einstein and the pure Gauss--Bonnet
equations,
\begin{eqnarray}
&&\epsilon _{abcdf}\,R^{ab}\wedge e^{c}\wedge e^{d}=0~,  \label{eq1mf} \\
&&\epsilon _{abcdf}\,R^{ab}\wedge R^{cd}=0~,  \label{eq2mf}
\end{eqnarray}%
which is certainly a severe restriction on the geometry, and not just a
correction to General Relativity. In particular, it is simple to see that
the only spherically symmetric solution satisfying both equations is flat
spacetime (see \emph{e. g.}, \cite{GOT}), and hence the five-dimensional
Schwarzschild geometry is ruled out as a solution.

It is amusing to see that requiring the metric to solve equations (\ref%
{eq1mf}) and (\ref{eq2mf}) simultaneously, is too strong so as to rule out a
spherically symmetric black hole, but not a $pp$-wave. Indeed, it is simple
to see that $pp$-wave solutions of the Einstein equations solve
independently (\ref{eq2mf}) as well as any equation constructed from powers
of the Riemann tensor with two free indices. This is because the Riemann
tensor for a $pp$-wave is orthogonal to a (covariantly constant) null vector
on all its indices, so that any term involving contractions of the Riemann
tensor, being quadratic or of higher degree identically vanishes. This is
the underlying reason of why $pp$-waves solve \textquotedblleft stringy
corrections" to General Relativity which involve higher powers of the
curvature to all orders (see \emph{e. g.}, \cite{HS}).

Even though the field equations turn out to be too restrictive for the
metric, they allow certain freedom when matter fields are switched on. The
purpose of the next subsection is to show that there exist configurations
corresponding to a brane-world with positive cosmological constant on the
worldsheet, supported by matter fields which are discontinuous, but not
singular, across the brane.

\subsection{Four--dimensional brane world solution}

Switching on the bosonic field $\kappa ^{ab}$, the field equations in the
torsionless sector read
\begin{eqnarray}
&&\epsilon _{abcdf}\,R^{ab}\wedge e^{c}\wedge e^{d}=-\epsilon
_{abcdf}\,R^{ab}\wedge D\kappa ^{cd}~,  \label{eq1k} \\
&&\epsilon _{\lbrack a|cdfg}\,R^{cd}\wedge \kappa ^{fg}\wedge e_{|b]}=0~,
\label{eq2k} \\
&&\epsilon _{abcdf}\,R^{ab}\wedge R^{cd}=0~.  \label{eq3k}
\end{eqnarray}%
Let us consider a domain wall of the form%
\[
ds^{2}=e^{2f(|z|)}\,\left( dz^{2}+\tilde{g}_{\mu \nu }^{(4)}dx^{\mu }dx^{\nu
}\right) ~,
\]%
where $\tilde{g}_{\mu \nu }^{(4)}=\tilde{g}_{\mu \nu }^{(4)}(x)$ is the
metric on the worldsheet. The vielbein can be chosen as%
\begin{equation}
e^{m}=e^{f(|z|)}\,\tilde{e}^{m}~,\qquad e^{4}=e^{f(|z|)}\,dz~,  \label{ans1}
\end{equation}%
where $\tilde{e}^{m}=\tilde{e}^{m}(x)$, with $m=0,\cdots ,3$, is the
vielbein along the worldsheet. The only non vanishing component of the
bosonic field $\kappa ^{ab}$ is assumed to be of the form%
\begin{equation}
\kappa ^{m4}=g(z)\,\tilde{e}^{m}~.  \label{ans2}
\end{equation}%
It is easy to see that the field equations (\ref{eq1k}-\ref{eq3k}) are
solved provided
\begin{equation}
\tilde{R}^{mn}=(f^{\prime }(|z|))^{2}\,\tilde{e}^{m}\wedge \tilde{e}%
^{n}~,\qquad e^{-2f(|z|)}f^{\prime }(|z|)=\frac{1}{2g(z)}~,  \label{presol}
\end{equation}%
where $\tilde{R}^{mn}$ is the curvature two-form along the worldsheet. This
means that $(f^{\prime }(|z|))^{2}$ must be a positive constant, which in
turn implies that $f(|z|)=-\xi |z|$. Thus, the geometry of the domain wall
acquires the form%
\begin{equation}
ds^{2}=e^{-2\xi |z|}\,\left( dz^{2}+\tilde{g}_{\mu \nu }^{(4)}dx^{\mu
}dx^{\nu }\right) ~,  \label{sol5d}
\end{equation}%
where the worldsheet metric of the four-dimensional brane-world must be
locally a de Sitter spacetime with radius $\ell =\xi ^{-1}$, and moreover,
the non-vanishing components of the bosonic field read
\begin{equation}
\kappa ^{m4}=-\frac{1}{2\xi }\mbox{sgn}(z)\,e^{-2\xi |z|}\,\tilde{e}^{m}~.
\label{solkappa}
\end{equation}

\section{Summary and discussion}

In this note, it is shown that deforming the Chern-Simons theory for AdS
gravity according to the expansion procedure of \cite{AIPV} is not sufficient
to produce a direct link with standard General Relativity. The fact that the
Einstein-Hilbert term appears in the action is just a mirage, since actually
the nonminimally coupled matter fields cannot be regarded as small corrections.
In fact, the dynamics suffers a radical departure from General Relativity so
that it is not continuously connected to the one obtained from the
Einstein--Hilbert action. Indeed, for vacuum configurations without torsion,
the field equations give severe restrictions on the geometry, so that in
particular, the metric must simultaneously satisfy the Einstein as well as the
pure Lovelock field equations. This precludes the existence of spherically
symmetric black holes, but does not rule out $pp$-waves.

In the five-dimensional case, it is shown that configurations corresponding to
a brane-world with positive cosmological constant on the worldsheet are
admissible when one of the matter fields is switched on. Curiously, the metric
for this class of solutions also solves the field equations of the supergravity
theories with local Poincar\'{e} invariance in vacuum\footnote{Note that this
class of solutions possesses a jump in the extrinsic curvature without the need
of a thin shell as matter source. This can be seen to be allowed by the
generalization of the Israel junction conditions for this kind of theories
\cite{Steve}.}, as discussed in \cite{HTZ}.

It goes without saying that the procedure can also be applied for the locally
supersymmetric extensions of these theories; however, it is clear that
supersymmetry would not improve the situation.

One can notice that there is an arbitrary element in the identification of the
gauge fields with spacetime geometry. For instance, in the five dimensional
case the connection (\ref{A5}) has two possible candidates to be identified
with the vielbein, namely, the fields $e^{a}$ and $h^{a}$, since both transform
as vectors under local Lorentz transformations. Choosing $e^{a}$, makes the
Einstein-Hilbert term to appear in the action, but as discussed here, this
choice brings in, apart from the Einstein equations (\ref{eq1mf}), the
Gauss-Bonnet ones (\ref{eq2mf}) to be simultaneously fulfilled by the geometry
in vacuum. Alternatively, if one identifies $h^{a}$ as the vielbein, the
Einstein-Hilbert term does not appear in the action, but the theory in vacuum
is well defined in the sense that the geometry must satisfy only the
Gauss-Bonnet equations, and the metric is not overdetermined as for the other
choice. The asymmetry between these two choices comes from the fact that in the
first case, the corresponding generators $P_{a}$ are actually pseudo
translations since $[P_{a},P_{b}]=Z_{ab}$, which brings in the extra field
equations to be required in vacuum. For the other choice, instead, the
corresponding generators $Z_{a}$ commute among themselves and can be identified
as the translation part of the Poincar\'{e} group; so, when matter fields are
switched off, there are no extra field equations for the geometry. Thus, for
this choice the nonminimally coupled matter fields actually correspond to
corrections that can be consistently switched off, but for the Gauss--Bonnet
theory.

From the last remark one can observe that in general, the expansion method
generates many fields that are candidates to be identified with the vielbein,
and for each choice a completely different gravitational sector arises. An
arbitrary choice is going to suffer of the same overdeterminacy of the
spacetime metric as it occurs for the choice that makes the Einstein-Hilbert
term to appear in the action. However, there is always one special choice that
is free from this problem. For this choice, the gravitational sector is
described by a lagrangian that is the dimensional continuation of the Euler
density of one dimension below. This is just the CS form for the Poincar\'{e}
group.

The expansion methods described here are also useful to connect an
eleven-dimensional AdS supergravity theory with a gauge theory for the
M-algebra \cite{HTZ2}.

\section*{Acknowledgments}

JDE is pleased to thank J. M. Izquierdo for interesting discussions and
comments. This work has been partially supported by grants 1051084, 1060831,
1040921, 1051056, 1061291 from FONDECYT. The work of JDE has been supported in
part by ANPCyT under grant PICT 2002 03-11624, by the FCT grant
POCTI/FNU/38004/2001, by MCyT, FEDER and Xunta de Galicia under grant
FPA2005-00188, and by the EC Commission under grants HPRN-CT-2002-00325 and
MRTN-CT-2004-005104, and by grant CONICYT/SECYT CH/PA03-EIII/014. Institutional
support to the Centro de Estudios Cient\'{\i}ficos (CECS) from Empresas CMPC is
gratefully acknowledged. CECS is funded in part by grants from the Millennium
Science Initiative,  Fundaci\'{o}n Andes and the Tinker Foundation.

\section*{Appendix. Extension to arbitrary odd dimensions}

As explained in the five-dimensional case, the expansion method of \cite{AIPV}
can be implemented for gravity as a Chern-Simons theory for the AdS group in
higher odd dimensions, through a similar shortcut consisting on shifting the
fields as in Eqs. (\ref{spin}), (\ref{vielbein}). The deformed action is then
obtained by replacing the shifted fields in the original action. For dimensions
$d=2n+1$, the generalized shift reads
\begin{equation}
\omega ^{ab} \rightarrow \omega ^{ab}+\sum_{k=1}^{n-1}\ell ^{-2k}\;\kappa
_{(k)}^{ab} ~, \qquad e^{a} \rightarrow \ell ^{-1}\;e^{a}+\sum_{k=1}^{n-1}\ell
^{-(2k+1)}\;h_{(k)}^{a}~,  \label{generalized shift}
\end{equation}%
where $h_{(k)}^{a}$, and $\kappa _{(k)}^{ab}$ are extra bosonic one-form
fields. The gravitational action, instead, is given by
\begin{equation}
I_{AdS}=\int \sum_{p=0}^{n}\frac{1}{(2n+1-2p)}{\binom{n}{p}}~\mathcal{L}%
^{(p)}  \label{LagAds}
\end{equation}%
where $\mathcal{L}^{(p)}$ are the dimensional continuation of the Euler forms
\begin{equation}
\mathcal{L}^{(p)}=\epsilon _{a_{1}\cdots a_{2n+1}}R^{a_{1}a_{2}}\wedge \cdots
\wedge R^{a_{2p-1}a_{2p}}\wedge e^{a_{2p+1}}\wedge \cdots \wedge e^{a_{2n+1}}
~.  \label{Lagp}
\end{equation}%
The deformed action is obtained after substituting (\ref{generalized shift}) in
(\ref{LagAds}) and retaining the terms of order $\ell ^{2-d}$. Its explicit
form is not particularly illuminating, but, as expected, contains the
Einstein--Hilbert term
\begin{eqnarray}
I^{2n+1} &=&\int \,\sum_{p=2}^{n}\frac{{\binom{n}{%
p}}}{(2n+1-2p)}\epsilon _{a_{1}\cdots a_{2p}a_{2p+1}\cdots
a_{2n+1}}\sum_{k=0}^{p}\sum_{k_{1}=0}^{k}\cdots
\sum_{k_{(n-2)}=0}^{k_{(n-3)}}\mathcal{C}_{p,k,k_{1},\cdots ,k_{(n-2)}}
\nonumber \\
&&\!\!\!\!\!\!\!\!\!\!\!\!\!\!R^{k_{(n-2)}}\left[
\prod_{j=1}^{n-2}R_{(j)}^{k_{(n-2-j)}-k_{(n-1-j)}}\right] R_{(n-1)}^{p-k}%
\sum_{r=0}^{2n+1-2p}\sum_{r_{1}=0}^{r}\cdots \sum_{r_{(n-2)}=0}^{r_{(n-3)}}%
\mathcal{C}_{2n+1-2p,r,r_{1},\cdots ,r_{(n-2)}}  \nonumber \\
&&\!\!\!\!\!\!\!\!\!\!\!\!\!\!e^{r_{(n-2)}}\left[
\prod_{i=1}^{n-2}h_{(i)}^{r_{(n-2-i)}-r_{(n-1-j)}}\right]
h_{(n-1)}^{2n+1-2p-r}+\frac{n}{2n-1}\epsilon _{a_{1}\cdots
a_{2n+1}}R^{a_{1}a_{2}}e^{a_{3}}\cdots e^{a_{2n+1}} ~.  \label{action2n+1}
\end{eqnarray}%
Here $R_{(j)}^{q}$ means the wedge product of $q$ curvatures defined by%
\[
R_{(k)}^{ab}=\sum_{i,j=1(i+j=k)}^{(k-1)}D(\omega )\,\kappa _{(k)}^{ab}+\kappa
_{(i)c}^{a}\wedge \kappa _{(j)}^{cb}~.
\]%
with the appropriate Lorentz indices, \textit{i.e.}~ $%
R_{(j)}^{a_{s+1}a_{s+2}}\cdots R_{j}^{a_{2q-1-s}a_{2q-s}}$ and the same
convention is adopted for $h_{(j)}^{q}$ that is the wedge product of $q$
one-form fields $h_{(j)}$. \footnote{We have also introduced the quantity
$\mathcal{C}_{r,s,s_{1},\cdots ,s_{N}}\equiv
{\binom{r}{s}}{\binom{s}{s_{1}}}\cdots {\binom{s_{N-1}}{s_{N}}}$. The
parameters are restricted as follows: for each fixed $p$ such that $2\leq p\leq
n$,
\[
j\left( k_{(n-2-j)}-k_{(n-1-j)}\right) +i\left( r_{(n-2-i)}-r_{(n-1-j)}\right)
+(n-1)(2n-p-k-r+1)=p-1~,
\]%
where the integers $i$ and $j$ run over $1,\cdots ,(n-2)$ with $%
k_{(n-2)}\leq k_{(n-3)}\leq \cdots \leq k_{1}\leq k\leq p$ and $%
r_{(n-2)}\leq r_{(n-3)}\leq \cdots \leq r_{1}\leq r\leq (2n+1-2p)$.} Besides,
we omit the wedge product symbol.

However, this action defines a Chern-Simons theory for the connection%
\[
\mathcal{A}=e^{a}\,P_{a}+\frac{1}{2}\,\omega
^{ab}\,J_{ab}+\sum_{k=1}^{n-1}h_{(k)}^{a}\,Z_{a}^{(k)}+\frac{1}{2}%
\,\sum_{k=1}^{n-1}\kappa _{(k)}^{ab}\,Z_{ab}^{(k)}~,
\]%
where one needs to include the aditional generators $Z_{a}^{(i)}$, and $%
Z_{ab}^{(i)}$ with $i=1,\cdots ,n-1$. The resulting algebra has the
following commutation relations%
\begin{eqnarray}
&&\left[ P_{a},P_{b}\right] =Z_{ab}^{(1)} ~,\qquad \left[ J_{ab},P_{c}\right]
=P_{a}\,\eta _{bc}-P_{b}\,\eta _{ac} ~, \qquad \left[ J_{ab},J_{cd}\right] =
-J_{ac}\,\eta _{bd}+\cdots ~, \nonumber  \label{algebra2n+1} \\
&&\left[ Z_{a}^{(i)},Z_{b}^{(j)}\right] = Z_{ab}^{(i+j+1)} ~, \qquad \left[
Z_{ab}^{(i)},Z_{c}^{(j)}\right] = Z_{a}^{(i+j)}\,\eta_{bc}
-Z_{b}^{(i+j)}\,\eta _{ac} ~, \nonumber \\
&&\left[ Z_{ab}^{(i)},Z_{cd}^{(j)}\right] =-Z_{ac}^{(i+j)}+\cdots ~,
\qquad \left[ P_{a},Z_{b}^{(i)}\right] = Z_{ab}^{(i+1)} ~, \nonumber \\
&&\left[ Z_{ab}^{(i)},P_{c}\right] = Z_{a}^{(i)}\,\eta _{bc}-Z_{b}^{(i)}\,
\eta_{ac}=\left[ J_{ab},Z_{c}^{(i)}\right] ~, \qquad \left[
J_{ab},Z_{cd}^{(i)}\right] =-Z_{ac}^{(i)}\eta _{bd}+\cdots
\end{eqnarray}
and the nonvanishing components of the $(n+1)$th--rank invariant tensor are
given by
\begin{eqnarray}
&&\left\langle J_{a_{1}a_{2}},\cdots
,J_{a_{2n-1}a_{2n}},Z_{a_{2n+1}}^{(n-1)}\right\rangle =\frac{2^{n}}{n+1}%
\,\epsilon _{a_{1}\cdots a_{2n+1}}~,  \nonumber \\
&&\Big<J_{a_{1}a_{2}},\cdots
,J_{a_{2i-1}a_{2i}},Z_{a_{2i+1}a_{2i+2}}^{(j_{1})},\cdots
,Z_{a_{2n-1}a_{2n}}^{(j_{q})},Z_{a_{2n+1}}^{(p)}\Big>=\frac{2^{n}}{n+1}%
\,\epsilon _{a_{1}\cdots a_{2n+1}}~,  \nonumber \\
&&\Big<J_{a_{1}a_{2}},\cdots
,J_{a_{2k-1}a_{2k}},Z_{a_{2k+1}a_{2k+2}}^{(l_{1})},\cdots
,Z_{a_{2n-1}a_{2n}}^{(l_{r})},P_{a_{2n+1}}\Big>=\frac{2^{n}}{n+1}\,\epsilon
_{a_{1}\cdots a_{2n+1}}~.  \nonumber  \label{brackets2n+1}
\end{eqnarray}
Here the numbers of generators $i$ and $q$ (resp. $k$ and $r$) satisfy $%
i+q=n $ (resp. $k+r=n$) and the other parameters are chosen such that $%
2(j_{1}+\cdots j_{q})+2(p-n+1)=0$, (resp. $2(l_{1}+\cdots l_{r})+2(1-n)=0$).

In the torsionless sector of the theory and in absence of matter fields, the
field equations are given by
\begin{eqnarray*}
&&\epsilon _{a_{1}\cdots a_{2n+1}}R^{a_{1}a_{2}}\wedge \cdots \wedge
R^{a_{2n-1}a_{2n}}=0~, \\
&&\epsilon _{a_{1}\cdots a_{2n+1}}R^{a_{1}a_{2}}\wedge \cdots \wedge
R^{a_{2n-3}a_{2n-2}}\wedge e^{a_{2n-1}}\wedge e^{a_{2n}}=0~, \\
&&\vdots \\
&&\epsilon _{a_{1}\cdots a_{2n+1}}R^{a_{1}a_{2}}\wedge e^{a_{3}}\wedge \cdots
\wedge e^{a_{2n}}=0~,
\end{eqnarray*}
which means that the metric must satisfy simultaneously the Einstein as well as
all possible ``pure Lovelock'' equations. This is an even more severe
restriction on the geometry than in the five-dimensional case, and not just a
correction to General Relativity. Analogously, it is simple to see that the
only spherically symmetric solution satisfying all the equations is flat
spacetime (see \emph{e.g.}, \cite{GOT}), so that Schwarzschild solution is
ruled out. However, it is simple to see that $pp$-wave solutions of the
Einstein equations solve the whole system.

%%%%%%%%%%%%%%%%%%%%%%%%%%%%%%%%%%%%%%%%%%%%%%%%%%%%%%%%%%%%%%%%%%%%%%%%%%%%%%%%

%%%%%%%%%%%%%%%%%%%%%%%%%%%%%%%%%%%%%%%%%%%%%%%%%%%%%%%%%%%%%%%%%%%%%%%%%%%%%%%%

\end{document}